\documentclass[aps,prb,twocolumn,floatfix,showpacs,superscriptaddress,fleqn]{revtex4-1}
\usepackage{amssymb}

\usepackage{graphicx}
\usepackage{amsmath}
\usepackage{bm}
\usepackage{color}

\usepackage[firstpage]{draftwatermark}
\usepackage[breaklinks]{hyperref}
\usepackage[all]{hypcap}

\hypersetup{
    plainpages=false,
    bookmarks=false,         
    unicode=false,          
    pdfborder={0 0 0}
    pdftoolbar=true,        
    pdfmenubar=true,        
    pdffitwindow=false,     
    pdfstartview={FitH},    
    pdftitle={},    
    pdfauthor={},     
    pdfproducer={LNCMI-CNRS-UJF-UPS-INSA}, 
    pdfkeywords={High} {Magnetic} {Fields}, 
    pdfnewwindow=true,      
    linktoc=section,
    colorlinks=true,       
    linkcolor=blue,          
    citecolor=red,        
    filecolor=magenta,      
    urlcolor=blue           
}

\begin{document}

\title{Identification of 2H and 3R polytypes of MoS$_{2}$ layered crystals using photoluminescence spectroscopy}

\author{S. Anghel}
\affiliation{Institute of Applied Physics, Academiei Str. 5,
Chisinau, MD-2028, Republic of Moldova}
\affiliation{Ruhr-Universitat Bochum, Anorganische Chemie III,
D-44801 Bochum Germany} \email{anggell@gmail.com}

\author{Yu. Chumakov}
\affiliation{Institute of Applied Physics, Academiei Str. 5, Chisinau, MD-2028, Republic of Moldova}

\author{V. Kravtsov}
\affiliation{Institute of Applied Physics, Academiei Str. 5, Chisinau, MD-2028, Republic of Moldova}

\author{A. Mitioglu}
\affiliation{Institute of Applied Physics, Academiei Str. 5,
Chisinau, MD-2028, Republic of Moldova} \affiliation{LNCMI,
CNRS-UJF-UPS-INSA, Grenoble and Toulouse, France}

\author{P. Plochocka}
\affiliation{LNCMI, CNRS-UJF-UPS-INSA, Grenoble and Toulouse,
France}

\author{K. Sushkevich}
\affiliation{State University of Moldova, Mateevici Str. 60,
Chisinau, MD-2009, Republic of Moldova}

\author{G. Volodina}
\affiliation{Institute of Applied Physics, Academiei Str. 5, Chisinau, MD-2028, Republic of Moldova}

\author{A. Colev}
\affiliation{Institute of Applied Physics, Academiei Str. 5, Chisinau, MD-2028, Republic of Moldova}

\author{L. Kulyuk}
\affiliation{Institute of Applied Physics, Academiei Str. 5, Chisinau, MD-2028, Republic of Moldova}

\SetWatermarkText{FIRST DRAFT PLEASE DO NOT DISTRIBUTE}
\SetWatermarkLightness{0.9}
\SetWatermarkScale{0.25}

\date{\today}

\begin{abstract}
The excitonic radiative recombination of intercalated Cl$_{2}$ molecules for two different polytypes 2H-MoS$_{2}$ and
3R-MoS$_{2}$ layered crystals are presented. The structure of the excitonic emission is unique and provides a robust
experimental signature of crystal polytype investigated. This result is confirmed by X-ray diffraction analysis and DFT
electronic band structure calculations. Thus, the bound exciton emission provides a nondestructive fingerprint for the
reliable identification of the polytype of MoS$_{2}$ layered crystals.

\end{abstract}

\maketitle



\section{Introduction}
Transition metal dichalcogenides (TMDC) is an emerging class of materials with an extremely wide spectrum of potential
applications, ranging from optoelectronic devices, field effect transistors and solar cell convertors to more mundane
lubricants.\cite{Lee2010a,Braga2012,Splendiani2010,Puthussery2011} Single layers of TMDC were first obtained by
mechanical exfoliation.\cite{Novoselov2005a} Due to their truly 2D character single layer TMDC's have attracted
considerable attention as strong potential candidates for the next generation of electronic devices. However, for
future applications the quality of the atomically thin crystals is extremely important.

In order to achieve the required high quality, the control of the synthesis of single crystals is crucial. Among the
different methods used to obtain single crystal, chemical vapor transport (CVT) method is widely used with a view to
device fabrication. In this method, halogen molecules are used as a transport
agent.\cite{Bougouma2013,Tiong2000,Hu2005} When synthesizing MoS$_{2}$ it is possible to obtain hexagonal (2H), as well
as, rhombohedral (3R) polytype layered crystals which have quite different physical properties. For example, the 3R
polytype of MoS$_{2}$, due to the non-centrosymmetric structure present in this form, exhibits a valley polarization in
photoluminescence emission even for a bulk crystal.\cite{Suzuki2014} To date the electronic properties of 3R phase
remain largely unexplored. Clearly, the development of an experimental probe to distinguish the polytype ``\emph{in
situ}'' is essential, especially since traditional Raman spectroscopy appears to be incapable of distinguishing between
the 3R or 2H crystal polytype.

In this paper we show that the photoluminescence spectra of excitons bound to halogen
molecules\cite{Kulyuk2002,Kulyuk2003,Kulyuk2005} in the van der Waals gap of the crystals provide a unique fingerprint
for the crystal polytype. X-ray analysis of the crystal structure and density functional theory (DFT) calculations of
electronic band structure of the layer type models of MoS$_{2}$ have been performed to support our findings.


\section{Sample characterization}
The synthetic 2H and 3R-MoS$_{2}$ single crystals have been grown by the vapor transport method, using Mo and S as
starting materials. The MoCl$_{5}$ compound, which decomposes at high temperatures, was used as a source of Cl$_{2}$
molecules for the CVT. The starting materials were placed in evacuated sealed quartz ampoules which were slowly heated
up to the synthesis temperature of 1150$^{\circ}$C for two days and maintained under these conditions for two days
more. Subsequently, the ampoules with the polycrystalline material were placed in a two-zone tube furnace. The
temperature of the crystallization chamber was set at around 930$^{\circ}$C in the region of the crystals growth,
according to references [\onlinecite{DAmbra1985,Baglio1983}]. The two polytypes were obtained by choosing a different
temperature gradient and a different concentration of the transport agent, which seems to have an important influence
on the growth process. The ampoules were held inside the furnaces for a period of up to 6 days, after which they were
slowly cooled to room temperature.

The 2H- and 3R- polytypes were identified and structurally characterized by the single crystal X-ray method. The
polymorphic purity of the as synthesized 2H- and 3R-polytypes (bulk samples) was confirmed by comparing the calculated
and experimental X-ray powder diffraction patterns. The X-ray diffraction data were obtained at room temperature using
an \emph{Xcalibur} \emph{E} diffractometer. The data were collected and processed using the program CrysAlisPro and
were corrected for the Lorentz and polarization effects and absorption \cite{Clark1995}. The structure was refined by
the full matrix least squares method on \emph{F}$^2$ with anisotropic displacement parameters using the program
SHELXL.\cite{Sheldrick2008} The unit cell parameters for 2H-MoS$_{2}$ are \emph{a}=\emph{b}=3.1625(1){\AA},
\emph{c}=12.300(1){\AA} and for 3R-MoS$_{2}$ \emph{a}=\emph{b}=3.1607(7){\AA}, \emph{c}=18.344(9){\AA}, and the
corresponding atomic coordinates are listed in Table~\ref{Tab1}. These parameters are in good agreement with those
reported in reference [\onlinecite{Schoenfeld1983}] which allows us to conclude that the intercalation with Cl$_{2}$
molecules does not change the parameters of the crystal structure.

\begin{table}
\begin{center}
\setlength{\tabcolsep}{15pt}
    \begin{tabular}{ccccc}
\hline \hline
\multicolumn{5}{c}{2H-MoS$_{2}$} \\
\cline{1-5}
          & $\emph{X}$ & $\emph{y}$ & $\emph{Z}$ & \emph{U}$_{eq}$\\
\hline
     Mo   &     1/3    &    2/3   &   0.25   &       6(1)    \\
     S    &     1/3    &    2/3   &  6228(1) &       7(1)    \\
\hline
    \end{tabular}
\end{center}

\begin{center}
\setlength{\tabcolsep}{16pt}
   \begin{tabular}{ccccc}
\hline \hline
\multicolumn{5}{c}{3R-MoS$_{2}$} \\
\cline{1-5}
          & $\emph{X}$ & $\emph{y}$ & $\emph{Z}$ & \emph{U}$_{eq}$\\
\hline
      Mo  &     0    &    0   &     0    &   12(1)    \\
      S1  &     0    &    0   &  2490(3) &   11(1)    \\
      S2  &     0    &    0   &  4189(3) &    9(1)    \\
\hline
\end{tabular}
\caption{Atomic coordinates $×10^{4}$ and equivalent isotropic displacement parameters ({\AA}$^{2}\times10^{3}$).
$\emph{U}_{eq}$ is defined as one third of the trace of the orthogonalized U$_{ij}$ tensor.}\label{Tab1}
\end{center}
\end{table}

\section{Photoluminescence measurements}
The PL spectra, recorded using a standard lock-in technique, were excited by the second harmonic emission of a
cw-operating YAG: Nd ($\lambda$ = 532nm). The samples were placed in a closed cycle cryostat operating in the
temperature range $10 - 100$K and the PL emission was collected from the 001 plane of the crystal (parallel to the
\emph{c}-axis).

\begin{figure}[]
\begin{center}
\includegraphics[width= 7.0cm]{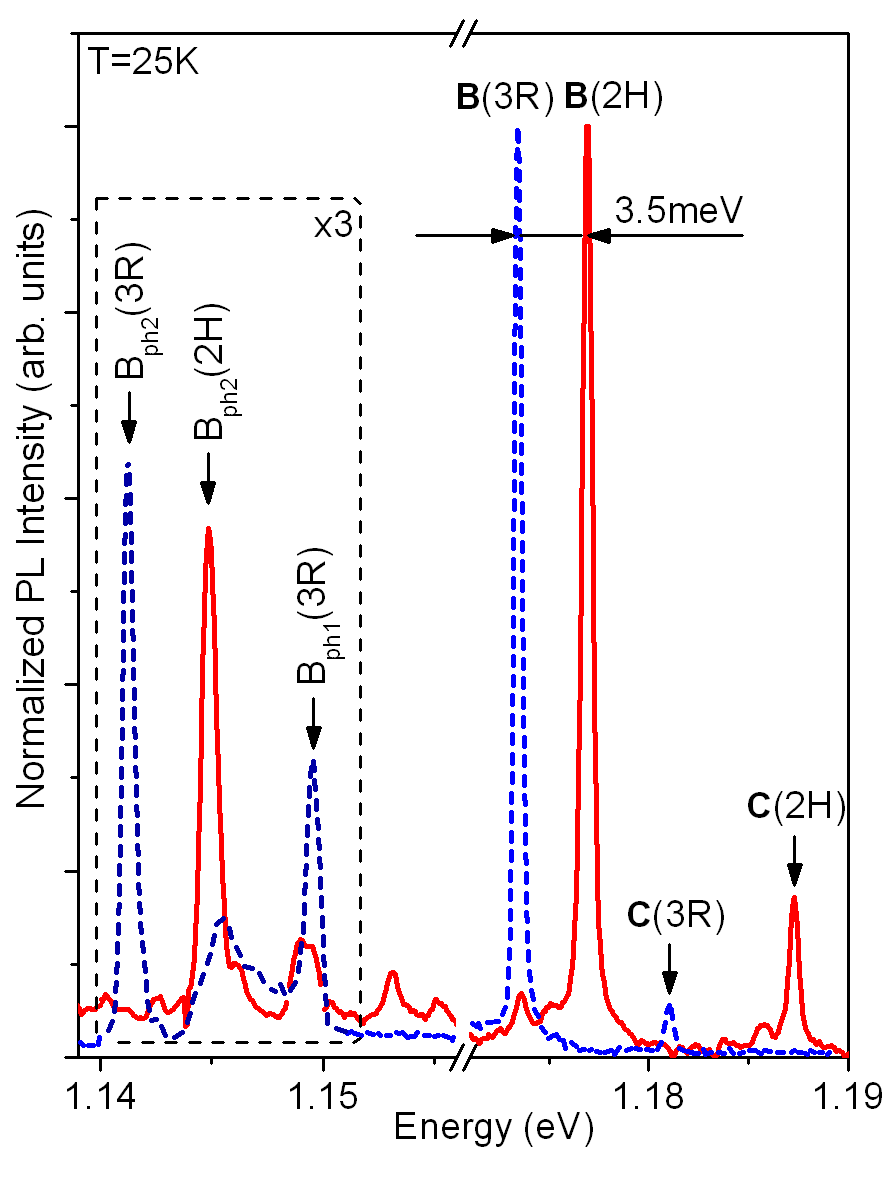}
\end{center}
\caption{(color online) Luminescent spectra at low temperature of Cl$_{2}$ intercalated 2H (blue broken line) and 3R
(red solid line) MoS$_{2}$ single crystals. }\label{Fig1}
\end{figure}

\begin{figure*}[]
\begin{center}
\includegraphics[width= 12cm]{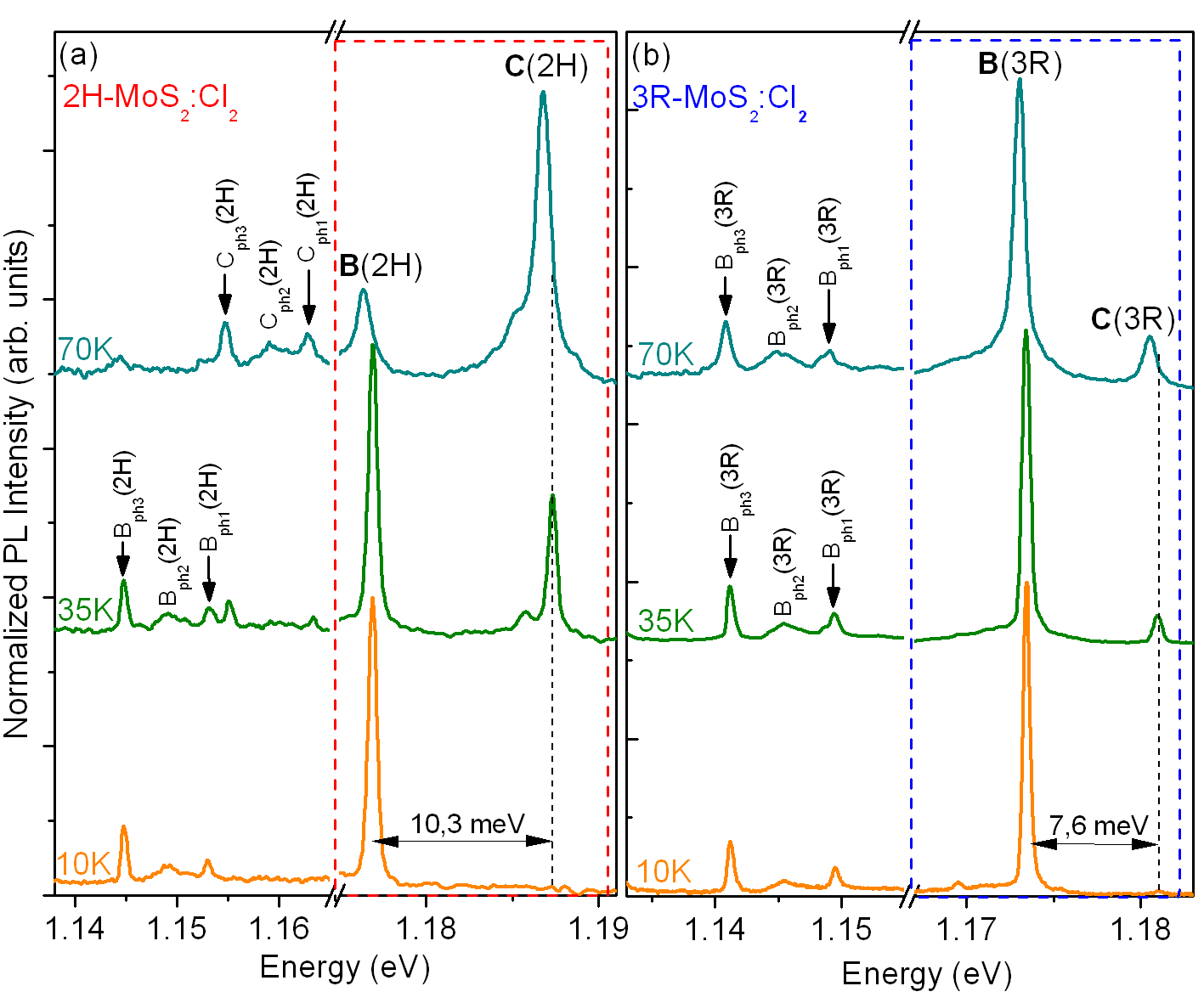}
\end{center}
\caption{(color online) PL spectra at different temperatures of (a) 2H-MoS$_{2}$ and (b) 3R-MoS$_{2}$ polytypes
measured at different temperatures. The dashed vertical lines are a guide to the eye indicating the position of the
weak $C$ excitonic lines at $10$K (which are nevertheless clearly visible in the $T=35$~K spectra).}\label{Fig2}
\end{figure*}

Representative PL spectra measured on the 3R and 2H polytypes of MoS$_{2}$ single crystals at $T=25$K are presented in
Fig.~\ref{Fig1}. The emission spectra of 2H and 3R polytypes are marked by red solid and blue broken lines
respectively. Each spectrum exhibits several sharp lines. The two strong lines in the energy range $1.17-1.19$~eV
correspond to the two zero-phonon excitonic lines. We label them $B$ and $C$ following our previous
notation.\cite{Kulyuk2003} 
The lower energy part of the spectrum is dominated by their phonon replicas where ${ph1}$-${ph3}$ symbols refer to the
number of the phonon replica. An analysis of the phonon replica sideband lines leads to the following three values of
phonon energy modes: E$_{ph1}$=23.9meV, E$_{ph2}$=28meV, and E$_{ph3}$=32.1meV. To the best of our knowledge this is
the first observation of halogen bound exciton emission in the case of the 3R-polytype. Fig.~\ref{Fig2} shows an
expanded view of PL spectra of the two polytypes measured at different temperatures. Although, the structure of the
spectrum for both polytypes is clearly similar, the emission the same excitonic complexes appears at significantly
different energies. For example, the 3R emission occurs at an energy 3.5~meV (for $B$ exciton) and 6~meV (for $C$
exciton) lower than for 2H emission. The resulting energetic separation of the excitonic lines is 10.3 meV for the 3R
crystal and only 7.6~meV for the 2H crystal. Thus, the size of the splitting provides a unique fingerprint for each
polytype which also has the advantage that it does not depend on the absolute calibration of the spectrometer.

Fig.~\ref{Fig2}(a),(b) also shows that the 2H and 3R emission evolves in a qualitatively similar manner with
temperature. For both polytypes the emission of exciton $C$ gains in intensity with respect to the emission from
exciton $B$. Quantitatively there are some differences. In contrast to 3R for which the exciton $B$ emission dominates
at all temperatures, in the 2H polytype the intensity ratio between $B$ and $C$ excitonic lines changes in favor of
exciton $C$ with increasing temperature. For both polytypes, increasing the temperature above $60$K leads to a rapid
quenching of all PL emission. Such a of the excitonic spectral lines in the case of 2H-MoS$_{2}$:Cl$_{2}$ crystals can
be understood in terms of non-radiative transitions (phonon emission) which dominate over radiative recombination
(photon emission) at higher temperatures.\cite{Colev2009,Kulyuk2003} A common characteristic of the emission spectra
for the both polytypes is that the excitonic region, very prominent at low temperatures, is accompanied by strong
phonon replicas: $C$$_{ph1}$-$C$$_{ph3}$ and $B$$_{ph1}$-$B$$_{ph3}$ (see Fig.~\ref{Fig2}(a),(b)). The energies of the
phonon replica emission are in good agreement with previous studies, \cite{Kulyuk2003} where they were interpreted as
local phonon modes induced by the center at the origin of the exciton related luminescence.

\section{Model description}
In order to explain the difference between 2H and 3R polytypes, we propose a model which describes the intercalation of
halogen molecules in the two polytypes. As the X-ray single crystal study did not reveal any essential changes in the
unit cell parameters in both studied polytypes, we have assumed the Cl$_{2}$ molecule is present in a relatively low
concentration. In other words, we consider that the halogen molecules disturb the crystal lattice only locally. The
intercalation of a considerable quantity of molecules should lead to a larger interlayer distance, compared to that in
the pure phase; indeed such an increase of the inter-layers separation upon intercalation has been
observed.\cite{Benavente2002,DAmbra1985} Due to the lack of structural information concerning the position of the
chlorine molecule, the DFT band structure calculations were carried out for the surfaces of 2H-MoS$_{2}$ and
3R-MoS$_{2}$ crystals. The studied surfaces were modeled by three molecular layers of polytypes that are without and
with intercalation of Cl$_{2}$ molecules (model I and II respectively). For model I, three layers and corresponding
atomic coordinates were taken from the structure of bulk crystals. In model II we assume that Cl$_{2}$ molecules are
intercalated in the van der Waals gap only between two layers and move them apart while the other interlayer gap
remains unchanged.

\begin{figure}[]
\begin{center}
\includegraphics[width= 6.5cm]{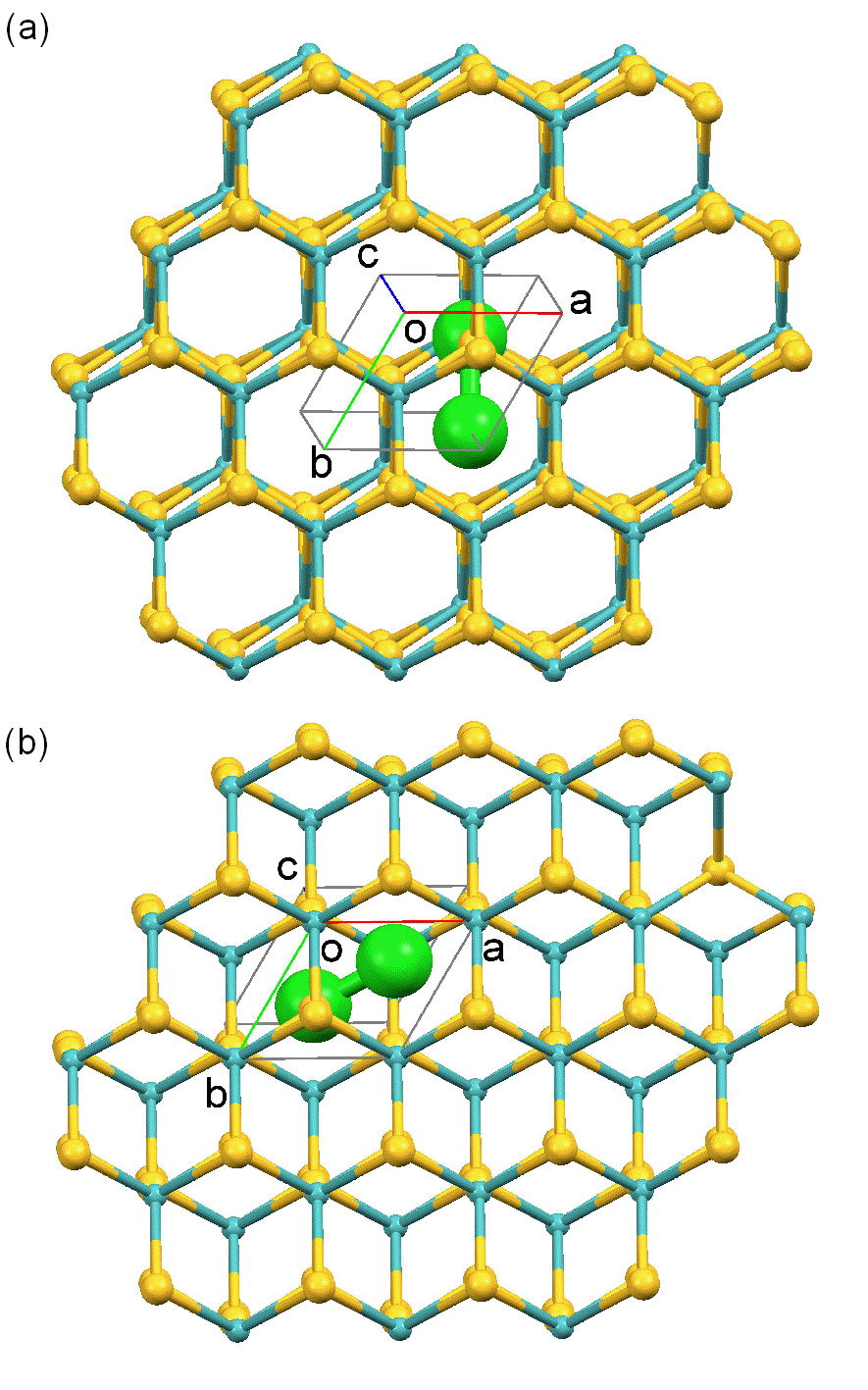}
\end{center}
\caption{(color online) (a), (b) Mutual arrangement of two
neibouring layers in the structure of 2H- (a) and 3R- (b) polytypes
and position of Cl$_{2}$ molecule between these layers. View
approximately along \emph{c} crystallographic axis. Mo, S and Cl
atoms are shown as spheres of arbitrary radii of cyan, yellow and
green color, respectively.}\label{Fig3}
\end{figure}

Possible positions of Cl$_{2}$ molecules intercalated between MoS$_{2}$ layers were chosen on the basis of the bulk
crystal structure analysis using PLATON tools.~\cite{Spek2003} The crystal packing of the layers revealed two types of
interstitial cavities in the interlayer of the van der Waals gap. These cavities are similar in both polytypes and are
formed by six or four surrounding sulphur atoms and have the shapes of a trigonal antiprism and a trigonal pyramid,
respectively. The coordinates of corresponding centroids of these cavities are 0, 0, 0.5 and 2/3, 1/3, 0.561 in
2H-polytype and 2/3, 1/3, 0.167 and 1/3, 2/3, 0.208 in 3R-polytype. The shortest distance between the centroids of two
such nearest cavities, which share a common face (three common sulpha atoms) equals 1.97{\AA} in both polytypes. On the
other hand a search of the Cambridge Structural Database (version 5.34) \cite{Bruno2002,Allen2002} and Inorganic
Crystal Structure Database \cite{Belsky2002} revealed that Cl - Cl distances in a chlorine molecule are between 1.96
and 1.98{\AA} and that the shortest Cl - S inter molecular contacts in the crystal are in the range 3.29-3.31{\AA}.
Thus, the centroids of conjugate cavities are complementary to the chlorine molecule and this position obviously
minimizes the necessary split of the interlayer of the van der Waals gap to adopt chlorine molecule taking into account
the van der Waals size of these molecules and minimal possible Cl - S intermolecular distance.

To place Cl$_{2}$ molecules in the desired positions, the gap between two MoS$_{2}$ layers in model II was extended to
provide the required Cl - S intermolecular distance. The separation between the planes of sulpha atoms from two
neighboring MoS$_{2}$ layers was increased up to 6.083 and 6.279{\AA} for the 2H- and 3R- polytypes, respectively,
compared with the corresponding separation of 3.021 and 2.997{\AA} in the bulky crystals and in model I. Although the
chlorine molecule has a similar nearest neighbor surrounding, the further environment in polytypes differs due to
unlike mutual arrangement of MoS$_{2}$ layers (Fig.~\ref{Fig3}(a), (b)). Even such small differences in halogen
positions may affect the radiative properties of excitons bound to the halogen molecules, leading to distinct emission
spectra.

Self-consistent ground-state calculations were performed with the ABINIT code~\cite{Gonze2002} to obtain the detailed
electronic structures. Electronic calculations were performed in the generalized gradient approximation with the
Perdew-Burke-Ernzerhof exchange-correlation energy functional.\cite{Perdew1996}

Orbitals are expanded in plane waves up to a cut-off of 25 Hartrees. The pseudo potentials used in our work were
generated from the pseudo potentials of Troullier-Martins.\cite{Troullier1991} The slab-model approach was used to
construct the surface-induced bulk alignment of the crystals. In a slab model the super cell with the dimensions
$3\emph{a}\times3\emph{b}$ along the layers and parameter along c axis, which corresponds to three layers, were
selected and repeated by use of periodic boundary conditions. When used with periodic basis functions (e.g., plane
waves), the repetition is performed in three dimensions and a vacuum layer with the thickness of 10{\AA} is introduced
to isolate the slabs. Thus, the super cell always includes three MoS$_{2}$ layers for both 2H and 3R polytypes in
models I and II. Only one chlorine molecule was introduced per super cell to model the low concentration. The
$3\emph{a}\times3\emph{b}$ dimension along the layer was chosen to perform calculations in reasonable computation time
with available recourses.

\section{Discussion}
For bulk MoS$_{2}$, the electronic states near the Fermi level are dominated by Mo 4\emph{d} and S 3\emph{p} levels.
Specifically, the conduction band states at the K point on the Brillouin zone, are primarily composed of strongly
localized \emph{d} orbitals at Mo atom sites. At the K point, the occupied part of the \emph{d} band has dominant
$d_{xy}-d_{x}^{2}-y^{2}$ character whereas the unoccupied portion is dominated by $d^{2}_{z}$ character.
\cite{Ramasubramaniam2011} They have minimal interlayer coupling since Mo atoms are located in the middle of the S-Mo-S
unit cell. The valence band maximum (VBM) is located at the $\Gamma$ point while the conduction band minimum (CBM) is
located about halfway between $\Gamma$ and K points; the gap is thus indirect having a value of 1.28~eV. On the other
hand, states near the $\Gamma$ point originate from a linear combination of Mo $d^{2}_{z}$ orbitals and the S $p_{z}$
orbitals and are fairly delocalized and have an antibonding nature. They have strong interlayer coupling and their
energies depend sensitively on the layer thickness. As a consequence, increasing the separation between consecutive
MoS$_{2}$ layers leads to weaker layer-layer interaction and lowers the energy of the antibonding states which causes
the VBM to shifts downwards. Thus, in the limit of widely separated planes, i.e., monolayer MoS$_{2}$, the material
becomes a direct gap semiconductor with a gap of about 1.9eV at $300$K.\cite{Reshak2005} Moreover, the stacking effect
on the interlayer bonding was confirmed by the low energy diffraction study of MoS$_{2}$ single
crystals,\cite{Mrstik1977} which shows that the inter-plane distance between the Mo and S atomic planes within the
topmost layer shrink about 5$\%$ compared to its bulk value. To resume, the electronic states at the $\Gamma$ point are
strongly affected by the long-range interlayer Coulombic interactions. In our case, halogen molecules intercalated
within MoS$_{2}$ layers should influence the interlayer interaction especially at the $\Gamma$ point. It should be not
forgotten that the intercalation of any molecules between the layers leads to an enlargement of the adjacent layers, as
was revealed in earlier publications.\cite{DAmbra1985,Benavente2002} This interaction is more pronounced in the case of
only few MoS$_{2}$ layers as was suggested in.\cite{Ellis2011}

\begin{figure*}[]
\begin{center}
\includegraphics[width= 12cm]{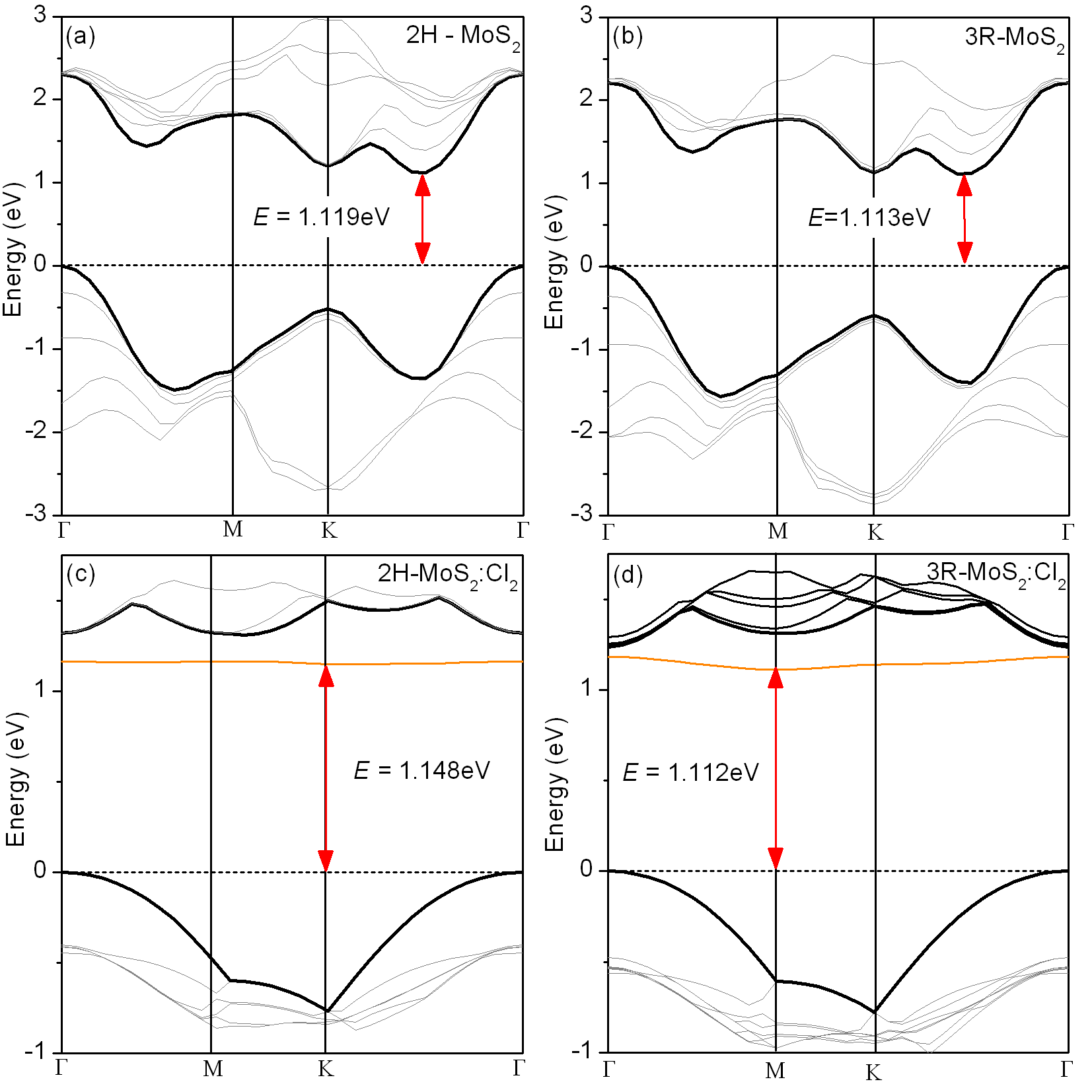}
\end{center}
\caption{(color online) Calculated electronic band structures of two
models containing three layers of MoS$_{2}$. Band structures of 2H
(a) and 3R (b) polytypes of MoS$_{2}$ where layers are fixed at bulk
positions. Electronic band structure of halogen intercalated 2H (c)
and 3R (d) polytypes of MoS$_{2}$. The orange line in (c) and (d)
correspond to the halogen level within the band gap. The coordinates
of \emph{k} points in the Brillouin Zones in studied super cells are
following: $\Gamma$: (0,0,0), M: (1/2,0,0), K: (1/3,1/3,0). The red
arrows indicate the minimum energy value. The horizontal dashed
lines indicate the valence band maximum.}\label{Fig4}
\end{figure*}

This assumption is confirmed by our theoretical calculations of the electronic structure of the two polytypes with and
without halogen intercalation, presented in Fig.~\ref{Fig4}. The model I has an indirect band gap structure, VBM being
located at the $\Gamma$ point while the CBM is located about halfway between $\Gamma$ and K points. The values of
indirect gaps are equal to 1.119 and 1.113 eV for 2H and 3R polytypes (Fig.~\ref{Fig4} (a),(b)); the values for the gap
of both polytypes are less than experimental ones but it is known that the density functional methodology
underestimates the band gaps of semiconductors compared to the experimental results (in our calculations, which are not
presented here, the E$_{g}$ for these bulk materials were equal to 1.062 and 1.019eV respectively, which are also
underestimated compared to experimental values). A similar band gap narrowing in 3R polytype, in comparison to that of
2H, was also observed in silicon carbide.\cite{Lambrecht1997}

Moreover, it was found for two single stacked sheets of MoS$_{2}$ that when the inter-sheet separation is greater than
4.5{\AA}, the band gap reaches the value found for a single sheet, as the inter-sheet interaction
vanishes.\cite{Li2007} For both polytypes in our first model the number of layers is the same but the interlayer
distances and stacking sequences are different. Within the stacking sequences ABAB for 2H-MoS$_{2}$ and ABCABC for
3R-MoS$_{2}$ the distances between the planes of sulphur atoms of neighboring S-Mo-S layers and distances between the
planes of sulpha atoms within the same layer are 3.021, 2.997 and 3.129, 3.117{\AA}, respectively. Moreover, in
3R-MoS$_{2}$ polytype the S-Mo-S layers are significantly shifted relatively each other in comparison with 2H-MoS$_{2}$
one (Fig.~\ref{Fig3}). Thus the structural features of these polytypes affect the long-range, interlayer Coulombic
interactions which led to a difference in values of the band gaps.

In the second model for two polytypes the Cl$_{2}$ intercalation creates energy levels in the band gap near the
conduction band edge which consist of 2\emph{p} antibonding states of Cl atoms while the valence bands are comprised of
\emph{d}-electron orbitals of Mo atoms. Moreover, halogen intercalation completely changes the local band structure and
the distribution of energy states over the Brillouin zone: both polytypes have now direct band gap transition at the
$\Gamma$ point with 1.323 and 1.246 eV values, for 2H and 3R polytypes, respectively (Fig.~\ref{Fig4} (c),(d)). The
values of halogen levels within the band gap of 2H and 3R polytypes are equal to 1.148 and 1.112 eV, respectively. The
intercalation of Cl$_{2}$ molecule has led to widening of the band gaps compared to the band structure of the polytypes
investigated in model I. The difference in the values of the given halogen levels is related to different positions of
Cl$_{2}$ molecules between the layers where they have different spatial surrounding. For 2H-MoS$_{2}$, only one Cl atom
in the molecule has a contact (4.32{\AA}) with Mo atom (Fig.~\ref{Fig3}(a)) that is less than above the indicted
threshold (4.5{\AA}) when band gap does not depend on inter-layer separation. In contrast for 3R polytype, the Cl atoms
in the molecule have both contacts with the metals and they are less than in 2H polytype and equal to 4.279 and
4.316{\AA}, respectively (Fig.~\ref{Fig3}(b)). Since in Cl$_{2}$ molecule the electrons are shared (not transferred),
there are no existing ions that are positive or negative charges, which leads to the creation of full outer shells of
chlorine atoms. This means that the repulsive forces hinder inter-layer coupling when Cl and Mo atoms are brought
closer to each other, leading to a widening of the band gaps of the polytypes. In turn, that leads to a greater
splitting of the antibonding states of Cl atoms in 3R-MoS$_{2}$, causing the lowering of the impurity levels in this
polytype in comparison with 2H polytype. The band gap widening for both polytypes in model II in comparison with model
I may be related to a long-range Coulombic interaction of chlorine molecules with the adjacent layers. For band
structure calculation, molecules intercalated within MoS$_{2}$ layers increase the distance between the two layers. In
order to understand the effect of halogen intercalation on the band structure transformation we calculated the energy
bands for model II with increased distances but without Cl$_{2}$ (not presented here). The obtained values are equal to
1.319 and 1.248 eV for 2H and 3R polytypes respectively. These values are very similar to direct band gap transition at
the $\Gamma$ point in model II for 2H, 3R-MoS$_{2}$ which means that the intercalation of chlorine molecules have
increased the distances between the adjacent layers which in turns have led to the local band gap widening.

Theoretical calculations have confirmed that the difference in halogen molecules positions in two polytypes should
result in different PL spectra separated by a small amount of energy which, indeed, is observed experimentally. The
temperature rise of the $C$ peak in the luminescent spectra to the detriment of the $B$ peak for 2H polytype occurs
because the level responsible for the $C$ emission peak has a radiative lifetime much shorter than the $B$
level.\cite{Colev2009} In the case of 3R polytype, despite a smaller (in comparison with 3R polytype) distance between
the $B$ and $C$ excitonic levels (7.5meV instead of 10.3meV), the temperature increase of the $C$-line relative
intensity is unimportant. This indicates that the radiative lifetimes of these levels are not significantly different,
and the relative intensity of lines $B$ and $C$ is determined mainly by their population (Boltzmann distribution).
However, this approach should be confirmed experimentally by kinetic (time resolved) measurements, which will be the
subject of a future publication.

\section{Conclusion}
To summarize, the synthetic MoS$_{2}$ single crystals were grown by means of the CVT method using Cl$_{2}$ molecules as
a transport agent. 2H-MoS$_{2}$ and 3R-MoS$_{2}$ polytypes were identified and structurally characterized using X-ray
diffraction to determine its structure under intercalation of Cl$_{2}$ molecules. The absence of any changes of the
unit cell parameters in both polytypes investigated indicates that the concentration of Cl$_{2}$ molecules in the
crystals is relatively low. Therefore, it was assumed that halogen molecules disturb the crystal lattice only locally
because their intercalation in large concentration would have led to a larger interlayer distance in comparison to that
in the pure phase. PL related to the excitons bound to the halogen molecules was investigated for the as grown
2H-MoS$_{2}$ and 3R-MoS$_{2}$ polytypes. It was shown that there is an evident difference between the low temperature
luminescence spectra of the investigated polytypes: Notably the excitonic splitting is significantly different
providing a robust signature of the polytype under investigation. The observed spectral shift and dissimilar behavior
of the spectra as a function of temperature are explained as consequence of slightly different positions of halogen
molecules in the interstitial space of the studied polytypes, which leads to a different interaction of the bound
excitons with the local crystal field. To interpret the obtained experimental results the DFT band structure
calculations were performed for three molecular layers of 2H-MoS$_{2}$ and 3R-MoS$_{2}$ polytypes without (model I) and
with (model II) intercalation of Cl$_{2}$ molecules. For model II, the structural features of these polytypes were
shown to affect the long-range interlayer Coulombic interactions, which led to a difference in values of the band gaps
in 2H-MoS$_{2}$ and 3R-MoS$_{2}$ due to slightly different positions of halogen molecules in the interstitial space of
these polytypes. The DFT band structure calculations for model II are in accordance with the spectroscopic experiments,
i.e., that the halogen bounded excitonic spectra of 2H polytype are shifted to higher energies compared to those of 3R
polytype. It was also shown that the halogen intercalation completely changes the local band structure and distribution
of energy states over the Brillouin zone of both polytypes.

\begin{acknowledgments}
We would like to thank Dr. Olga Iliasenco for technical assistance.
The investigation has been carried out in the framework of the
HP-SEE project (the fast track access to HPC cluster resources
available in the South-East Europe region) and of the STCU project
Nr. 5809 (financial support for a part of the research).
\end{acknowledgments}

\bibliography{3R2Hbib}

\end{document}